\documentclass{emulateapj}
\usepackage{graphicx,times,subfigure}
\usepackage{amsmath}
\usepackage{natbib}
\usepackage{xfrac}
\usepackage[T1]{fontenc}
\usepackage{aecompl}
\usepackage{amssymb}
\usepackage{hyperref}
\usepackage{multirow}

\hypersetup{
  colorlinks= true, 
  urlcolor  = Blue, 
  filecolor=black,
  runcolor=blue,
  linkcolor = Maroon, 
  citecolor = Mulberry 
} 

\usepackage[usenames, dvipsnames]{xcolor}

\newcommand{\vel}{\mathbf{u}}
\newcommand{\zhat}{\widehat{\mathbf{z}}}
\newcommand{\sol}{\mathbf{y}}
\newcommand{\lrb}[1]{\left[#1\right]}
\newcommand{\lr}[1]{\left(#1\right)}

\begin{document}
\title{Extending a Physics-Informed Machine Learning Network for Superresolution Studies of Rayleigh-B\'{e}nard  Convection }
\shorttitle{Superresolution Studies of Rayleigh-B\'{e}nard  Convection  }

\author{Diane M.~Salim\altaffilmark{1, 2},  Blakesley~Burkhart\altaffilmark{1, 2} \& David~Sondak\altaffilmark{3}\\
  }

\altaffiltext{1}{Department of Physics and Astronomy, Rutgers University,  136 Frelinghuysen Rd, Piscataway, NJ 08854, USA}
\altaffiltext{2}{Center for Computational Astrophysics, Flatiron Institute, 162 Fifth Avenue, New York, NY 10010, USA}
\altaffiltext{3}{Dassault Systemes Simulia Corp.}

\shortauthors{Salim, Burkhart and Sondak}

\begin{abstract}
Advancing our understanding of astrophysical turbulence is bottlenecked by the limited resolution of numerical simulations that may not fully sample scales in the inertial range. Machine learning (ML) techniques have demonstrated promise in up-scaling resolution in both image analysis and numerical simulations (i.e., superresolution). Here we employ and further develop a physics-constrained convolutional neural network (CNN) ML model called "MeshFreeFlowNet'' (MFFN) for superresolution studies of turbulent systems. The model is trained both on the simulation images as well as the evaluated PDEs, making it sensitive to the underlying physics of a particular fluid system.  We develop a framework for 2D turbulent Rayleigh-Bénard convection (RBC) generated with the \textsc{Dedalus} code by modifying the MFFN architecture to include the full set of simulation PDEs and the boundary conditions. Our training set includes fully developed turbulence sampling Rayleigh numbers ($Ra$) of $Ra=10^6-10^{10}$.  We evaluate the success of the learned simulations by comparing the power spectra of the direct \textsc{Dedalus} simulation to the predicted model output, and compare both ground truth and predicted power spectral inertial range scalings to theoretical predictions. We find that the updated network performs well at all $Ra$ studied here in recovering large-scale information, including the inertial range slopes. The superresolution prediction is overly dissipative at smaller scales than that of the inertial range in all cases, but the smaller-scales are better recovered in more turbulent, than laminar, regimes. This is likely because more turbulent systems have a rich variety of structures at many length scales compared to laminar flows.
\end{abstract}

\section{Introduction}\label{sec:intro}

Many astrophysical and geophysical fluid flows are governed by the phenomenon of thermal convection~\citep{lappa2009thermal} in which fluid motion is driven by a source of heat. Systems of scientific and engineering interest typically have parameter regimes that lead to turbulent flow fields. Turbulence is a ubiquitous nonlinear, multiscale phenomenon and is often considered one of the last remaining open problems in classical physics. The equations governing fluid flow have been studied for well over a century, but have remained analytically intractable in all but the simplest cases and a simple but general phenomonology of turbulence has proved elusive. With the rapid rise in computational resources, numerical simulations have proven to be an indispensable and crucial tool for complementing theory and experiments in studying turbulent flows and making accurate predictions of quantities of interest in systems involving turbulence \citep{Boldyrev2002,ChoLazarianVishniac2002,ChoLazarian2003,BeresnyakEtAl2005}. Moreover, the data generated by numerical simulations of turbulence has lead to creative and powerful data analysis techniques and given rise to some of the early data-driven modeling efforts \citep{2019AnRFM..51..357D,Peek2019,2020AnRFM..52..477B}. Nevertheless, simulations of turbulence face their own challenges, most notably that achieving the parameter regimes found in nature requires computational resources that are  not available now or for the foreseeable future \citep{Burkhart2021}. In light of these challenges, researchers have developed efficient models to study turbulent flows at reduced computational cost while retaining acceptable accuracy.

The spatial and temporal resolution of numerical simulations is limited, which restricts the scales that can be represented in a simulation. Turbulence is characterized by a high Reynolds number ($Re$), which represents the ratio of inertial to viscous forces. In a direct numerical simulation (DNS), all the scales of the turbulent flow are fully resolved and the equations of motion are solved to machine precision. The numerical resolution of DNS scales roughly with $Re^{3}$~\citep{pope2000turbulent} or worse~\citep{buaria2019extreme, buaria2022vorticity}. It is not uncommon for astrophysical and geophysical flows to have $Re > 10^{9}$, which results in a cost-prohibitive DNS. To overcome this issue, researchers have developed many creative models including reduced order models~\citep{berkooz1993proper,schmid2010dynamic} and turbulence models~\citep{pope2000turbulent, sagaut2005large, pouquet1976strong}. In recent years, the emergence of ML has given rise to a resurgence in the development of data-driven models to augment and enhance the current modeling paradigm.

Simulations of turbulence have led to the development of numerous algorithms and sophisticated statistical diagnostic techniques \citep{Koch_2019,Bialy_2020,Saydjari_2021,Burkhart2021} for working with large datasets over the past half-century. Over the past two decades, new computing paradigms and novel ML algorithms have been applied to turbulence datasets in an attempt to shed light on turbulence~\citep{buaria2022vorticity, ghazijahani2022benefits, pandey2020perspective}, develop new turbulence models~\citep{ling2016reynolds, bae2022scientific}, create novel reduced order models of turbulence and chaotic systems~\citep{vlachas2022multiscale, sondak2021learning, linot2020deep}, augment and enhance traditional numerical methods~\citep{kochkov2021machine}, and generate statistically realistic turbulence datasets and boundary and initial conditions without running large simulations~\citep{fukami2019super, Jiang2020, fukami2021machine, ren2022physics, kim2021unsupervised}. There has recently been substantial interest in the development and application of superresolution algorithms to turbulence datasets~\citep{ren2022physics}. Indeed, because turbulence datasets can be large and stress storage resources, it may be beneficial to save very low-resolution data and reconstruct DNS-level detail when the dataset is ready to be used. Another intriguing application could be the reconstruction of DNS results from large eddy simulations. This would circumvent the need to perform computationally intensive DNS runs while potentially delivering DNS quality results.

Progress at the intersection of turbulence and superresolution is rapidly evolving. Researchers have developed supervised models with features such as downsampled skip-connection/multiscale models \citep{fukami2019super, fukami2021machine} and PDE-solving capacities through  multilayer perceptron (MLP) branches and subsequent PDE losses, \citep{Jiang2020, ren2022physics}, as well as unsupervised~\citep{kim2021unsupervised} neural network (NN) architectures for superresolution of turbulent flows. In the present work, we build off of an architecture called MeshFreeFlowNet (MFFN), first presented in ~\citet{Jiang2020} (hereafter J20), and enhance that model by injecting additional physical constraints including boundary conditions and the divergence-free constraint on the velocity field. Additionally, we extend the Rayleigh-B\'{e}nard dataset used in that work, a system whose degree of turbulence is characterised by the Rayleigh number ($Ra$), which is the ratio of inertial driven buoyancy forces to diffusive forces. In this study we run simulations to higher $Ra$ than that presented in J20 and ensure that the training samples for each $Ra$ are from the statistically steady state and sampled over several eddy turnover times.

The governing equations and a review of the MeshFreeFlowNet architecture are presented in Section~\ref{sec:methods}. The main results are presented in Section~\ref{sec:results} followed by a discussion in Section~\ref{sec:discussion}. Conclusions are drawn and potential future avenues are discussed in Section~\ref{sec:conclusions}.

\section{Methodology} \label{sec:methods}

    
\subsection{Simulations of Rayleigh-B\'{e}nard Convection}
    
\begin{figure}
    \includegraphics[width=\linewidth]{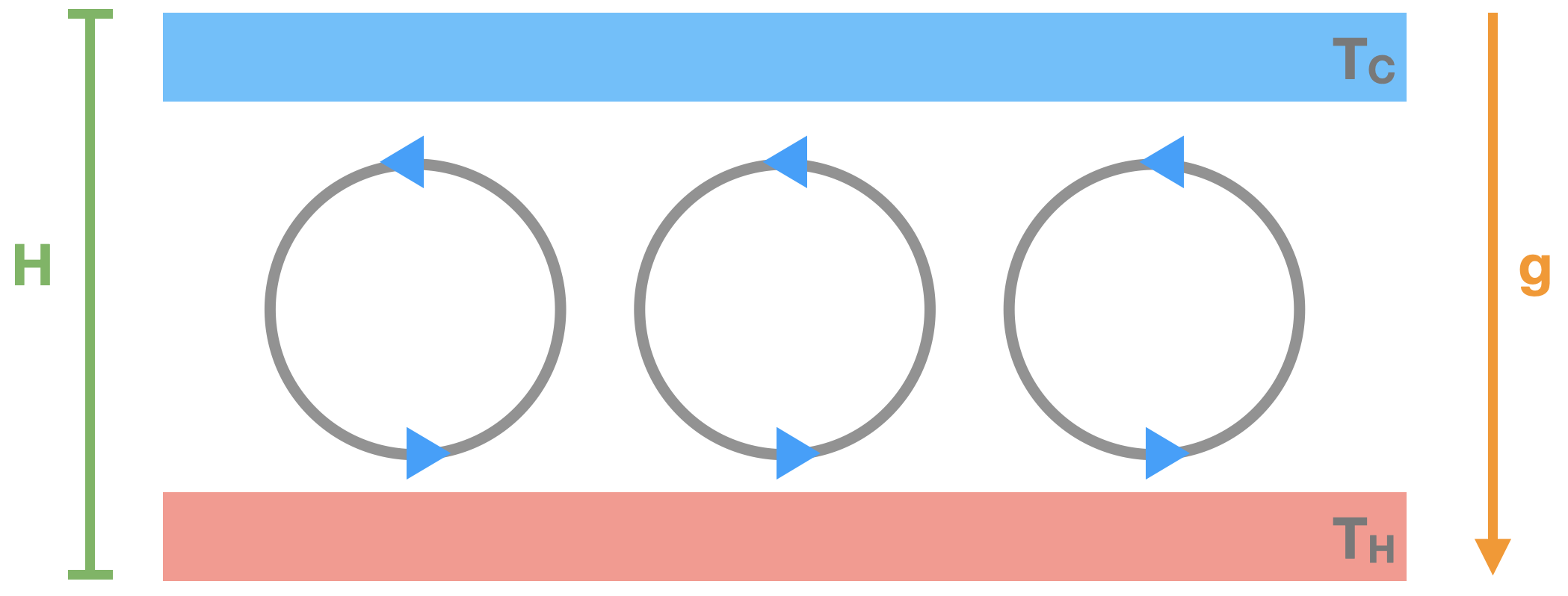}
    \caption{Schematic diagram showing the physical set-up that leads to RBC. Two parallel plates at temperatures $T_c$ and $T_H$ where $T_c<T_H$ are separated by a distance $H$ and are under the influence of a gravitational acceleration $g$.} 
\label{fig:RBC_diagram}
\end{figure} 

\textbf{A physics-informed NN \citep{RaissPerdikarisKarniadakis2019, CaiEtAl2021, Karniadakis2019, SongEtAl2019, KharazmiZhangKarniadakis2019, WangEtAl2020} } is explored for super-resolution of data generated from simulations of turbulent RBC. This physical system was selected because it is representative of many astrophysical and geophysical fluid scenarios. RBC is concerned with the buoyancy-driven flow of a fluid heated from below and cooled from above. The system is set up as a fluid under the influence of gravity and confined between two parallel plates of differing temperatures which are separated by a distance $H$ (see Figure~\ref{fig:RBC_diagram}). The Boussinesq approximation is used which assumes that density varies linearly with temperature in the buoyancy force and is otherwise negligible. The nondimensional Boussinesq equations are solved for the nondimensional velocity $\vel=\lr{u, v}$ and the nondimensional temperature $T$ in a two-dimensional domain with dimensionless $\mathbf{x}=\left(x, z\right)$ where $x\in\left[0, 1\right]$ and $z\in\left[-1/2, 1/2\right]$. The governing partial differential equations (PDEs) are 
\begin{align}
  \mathcal{R}_{M}\lrb{\sol} &= \dfrac{\partial \vel}{\partial t} + \vel\cdot\nabla\vel + \nabla P - \nu_{*}\nabla^{2}\vel - T\zhat = 0 \label{eq:mom} \\
  \mathcal{R}_{C}\lrb{\sol} &= \nabla\cdot\vel = 0 \label{eq:cont} \\
  \mathcal{R}_{E}\lrb{\sol} &= \dfrac{\partial T}{\partial t} + \vel\cdot\nabla T - \kappa_{*}\nabla^{2}T = 0 \label{eq:energy}
\end{align}

where 
$\mathcal{R}_{M}\lrb{\sol}$, $\mathcal{R}_{C}\lrb{\sol}$, and $\mathcal{R}_{E}\lrb{\sol}$ are the residuals for the momentum, continuity, and energy equations, respectively, $\sol = \lr{u, v, P, T}$ is the solution vector, and $\widehat{\mathbf{z}}$ is the unit vector in the wall-normal direction. Note that the solution residuals are identically zero for $\sol$ satisfying the governing equations. 
Equations~\eqref{eq:mom}-~\eqref{eq:energy} were obtained by nondimensionalizing time by the free-fall time $\tau = \sqrt{H/g^{\prime}}$, velocity by the free-fall speed $u_{\mathrm{ff}}=H/\tau$, pressure by the dynamic pressure $P_{dyn}=\rho_{0}u_{\mathrm{ff}}^{2}$, and temperature by half the temperature difference between the top and bottom plates $T_{\Delta} = \Delta T / 2$. The free-fall time involves the reduced gravity $g^{\prime} = g\alpha_{V}\Delta T /2$ where $g$ is the acceleration due to gravity and $\alpha_{V}$ is the coefficient of volumetric expansion. This nondimensionalization results in the two primary dimensionless parameters in equations~\eqref{eq:mom}-~\eqref{eq:energy},
\begin{align}
  \nu_{*} &= \dfrac{\nu}{\sqrt{g^{\prime}H^{3}}}\label{eq:nu_star} \\
  \kappa_{*} &= \dfrac{\kappa} {\sqrt{g^{\prime}H^{3}}}\label{eq:kappa_star}.
\end{align}
These dimensionless diffusivities are related to the classical dimensionless parameters via 
\begin{align}
  \nu_{*} &= \sqrt{\dfrac{16 Pr}{Ra}} \\
  \kappa_{*} &= \sqrt{\dfrac{16}{RaPr}}
\end{align}
where
\begin{align}
  Ra = \dfrac{g\alpha_{V}\Delta T H^{3}}{\nu\kappa}
\end{align}
is the Rayleigh number, which represents the ratio of buoyancy driven inertial forces to viscous forces and 
\begin{align}
  Pr = \dfrac{\nu}{\kappa}
\end{align}
is the Prandtl number, which is a fluid property representing the ratio of kinematic to thermal diffusivity. As a point of reference, $Ra$ is roughly related to the Reynolds number $Re$ via $Ra\sim Re^{1/2}$ for convective flows. Turbulent thermal convection involves extraordinarily large values of $Ra$ and a fundamental scientific question is concerned with the amount of heat transported in such turbulent systems. The Nusselt number ($Nu$) is the primary diagnostic dimensionless parameter in RBC. It represents the amount of heat transported from the bottom plate to the top plate. Given the non-dimensional formulation presented above, the unsteady $Nu$ is defined as 
\begin{align}
  Nu_{t}\lr{t} = -\left.\dfrac{d\overline{T}}{dz}\right|_{z_{\text{wall}}} \label{eq:Nusselt}
\end{align}
where $\overline{\lr{\cdot}}$ represents a spatial average over the plane parallel to the walls. In statistically steady state, after integrating over the width of the fluid layer, the $Nu$ can be written 
\begin{align}
  Nu = 1 + \dfrac{\left<vT\right>}{\kappa_{*}}. \label{eq:Nu_sss}
\end{align}
From~\eqref{eq:Nu_sss}, it is clear that in the purely conductive state (i.e. when $v=0$), $Nu=1$. Equations~\eqref{eq:mom}-~\eqref{eq:energy} use periodic boundary conditions in the $x$ direction for both velocity and temperature. No-slip velocity boundary conditions are used at the top and bottom walls. The temperature at the top wall is $T\left(z=1/2\right)=-1/2$ and the temperature at the bottom wall is $T\left(z=-1/2\right)=1/2$. The system of equations was solved numerically and the resulting fields were used as the training data for the ML algorithm. The unsteady $Nu$ was monitored to determine when the simulation reached a statistically steady state.  

\begin{figure}
    \includegraphics[width=\linewidth]{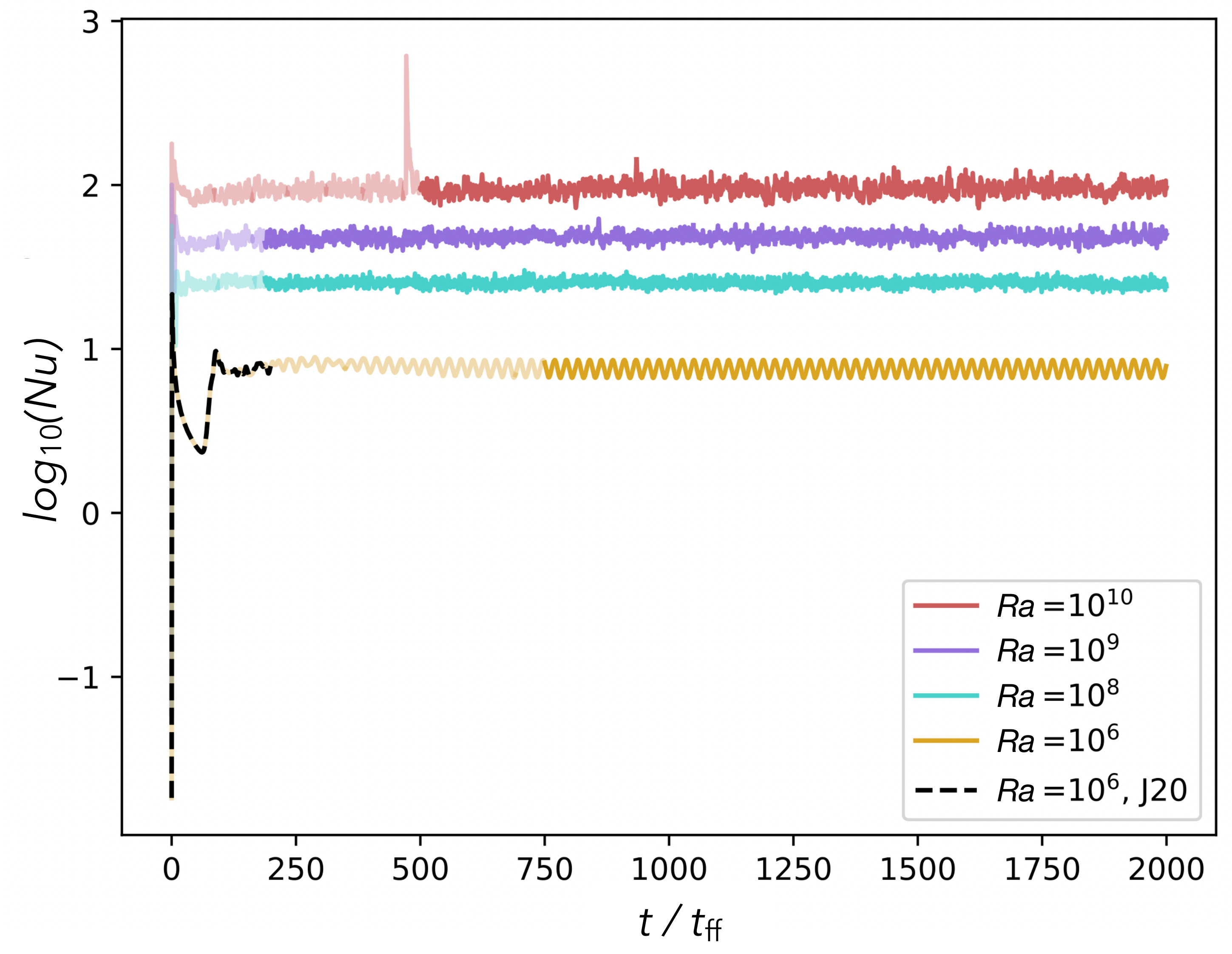}
    \caption{
    Temporal evolution of the Nusselt number ($Nu$) for the simulations created for our training set, each with differing Rayleigh number ($Ra$) values, with the $x$-axis showing the time in units non-dimensionalized by the free-fall time $t_{\mathrm{ff}}$. We limit our training set creation to originate from the subset in $t/t_{\mathrm{ff}}$ that we deem to be in statistically steady-state, indicated in bold. The origin of \citep{Jiang2020}'s training set which was derived from a simulation of $Ra=10^6$ is shown in the black dotted line.} 
\label{fig:nusselt}
\end{figure}

 

We generate the training data set for the ML pipeline using the spectral method code \textsc{Dedalus}~\citep{Burns2020}.  \textsc{Dedalus} solves general sets of partial differential equations using a spectral method approach and, for the purpose of this paper, we solve the RBC Equations~\eqref{eq:mom}-~\eqref{eq:energy}. 
The low $Ra$ simulations use a spatial resolution of $N_x=128$ and $N_z=512$ 
while the higher $Ra$ simulations use $N_x=256$ and $N_z=1024$. 

Our simulation setup is similar to J20, but we make several modifications. First, J20 had only considered an oscillatory RBC system in transition before a statistically steady-state was reached. In our work we seek to understand the performance of the super-resolution network in the turbulent steady state.  Second, J20 considered $Ra$ in the range of $Ra=10^4-10^8$, which is non-turbulent to mildly turbulent.  A primary aim of our study is to understand the performance of MeshFreeFlowNet at large $Ra$ (up to 10$^{10}$) which is in the highly-turbulent regime. Therefore in our study we run 4 separate simulations that span $Ra=10^6$ (non-turbulent) to $Ra=10^{10}$ (turbulent) until a statistically steady state is reached. Each simulation is run at a different $Ra$, namely $Ra=10^6, ~10^8, ~10^9$ and $10^{10}$, and each separate training run of MFFN is only trained on one $Ra$. $Pr=1$ is used in all simulations.

In Figure~\ref{fig:nusselt} we show the $Nu$ evolution of  our simulations (coloured lines). Each line corresponds to a seperate simulation. We overplot the J20 training set as a dashed black line for reference. The J20 simulation had only begun to reach a statistically steady state. In developing our training data, we use the $Nu$ evolution to  estimate the time after which statistically steady-state has been reached. This typically occurs after $250$ time steps.  From these subset we randomly sample low resolution ``slab'' of $128\times128$ pixels in the spatial dimension and $8$ time steps in the time dimension to use as training examples for each $Ra$. In total, 3000 slab samples are extracted for training and 2 samples for validation for \textit{each} of the 4 simulation runs. 

    
\subsection{Architectural backbone: MeshFreeFlowNet}
\begin{figure*}
    \includegraphics[width=\linewidth]{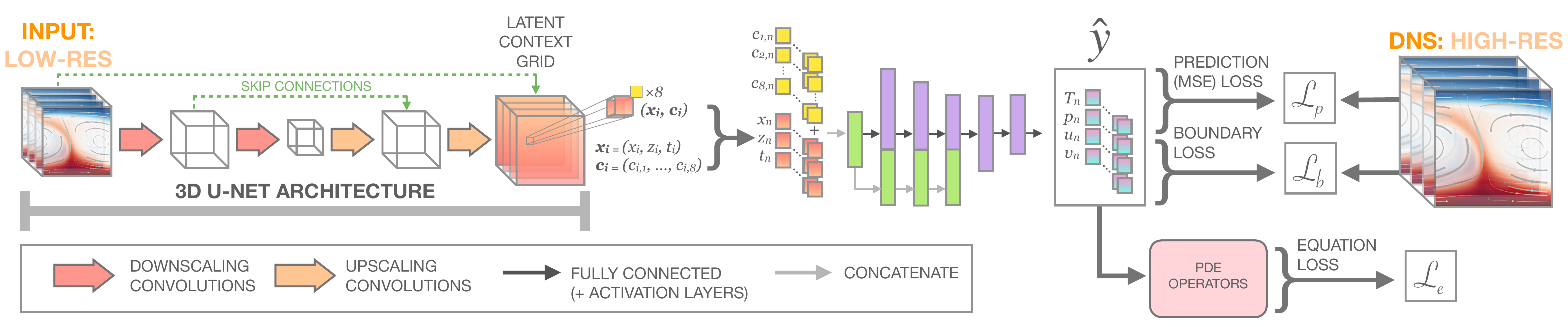}
    \caption{Schematic diagram of our modified MeshFreeFlowNet architecture. The components of the architecture shown in the top row of the diagram represent the Convolutional Neural Network (CNN) constituents of the model. Namely, a coarsened 2D direct numerical simulation (DNS) is passed through a U-Net structure \citep{Ronneberger2015}, resulting in a Latent Context grid (LCG; represented as the pink-orange array). In our visual representation of information flow, pink arrows indicate a series of downscaling convolutions whilst orange arrows that of upscaling convolutions. Latent context vectors are randomly sampled from $n=1024$ points in the LCG represented in the figure as the small cube being extracted from the LCG. Associated with each extracted point are latent context vectors at each corner of the grid-point, one of which is represented by the small yellow square in the corner of the extracted cube. These points are passed into a multilayer perceptron (MLP). The MLP generates the solution fields, the set of which is denoted as $\widehat{\mathbf{y}}$ and represented in the figure as the set of small pink-blue squares in the grey rectangle. These values are compared to the high resolution DNS and results in a traditional prediction loss ($\mathcal{L_P}$) as well as a boundary loss ($\mathcal{L_B}$). The outputs of the MLP are additionally put through partial differential equation (PDE) operators, represented as the light pink rounded box in the figure, to determine PDE residuals. The deviation of these residuals from $0$ indicate the degree to which the PDEs of the system are satisfied, resulting in an equation loss ($\mathcal{L_E}$).} \label{fig:architecture}
\end{figure*} 

The starting point for the present work is the ``MeshFreeFlowNet'' neural network architecture pipeline~\citep{Jiang2020}. The input to train MeshFreeFlowNet is the coarsened 2D temporal simulation slabs described above. MeshFreeFlowNet also uses the underlying partial differential equations and outputs the high-resolution counterpart of the coarse inputs. We now review the different steps that the architecture follows in generating the output.

The coarsened, low-resolution spatial and temporal data of the four simulated fields (temperature, pressure, and two components of velocity) is denoted by $\mathcal{D_L}$. $\mathcal{D_L}$ is passed through a 3-dimensional U-Net architecture \citep{Ronneberger2015}, $\Psi_{\theta_1}$, with $\theta_1$ being the set of learnable parameteres associated with the CNN. $\Psi_{\theta_1}$ is comprised of a contractive part and expansive part, both of which contain a residue block of 3 convolutional layers ($1\times1, ~3\times3, ~1\times1$), a batch normalization layer and a ReLU activation layer. Following the block in the contractive part is a maxpooling layer of stride 2, whereas in the expansive part it is followed by a layer performing nearest neighbour upsampling. 
In Figure~\ref{fig:architecture}, a series of downscaling convolutions representing contractive parts are denoted by pink arrows and series of upscaling convolutions representing the expansive parts by orange arrows. Furthermore, the layers in the contractive parts are concatenated to the layer of similar grid size in the expansive parts via a skip connection, which preserves localized contextual information. These skip connections are represented as the the dashed green lines in Figure~\ref{fig:architecture}. The  is the Latent Context Grid (LCG), $\mathcal{G}$, visualized as the pink-orange array in Figure~\ref{fig:architecture}. Thus: 
\begin{align}
    \mathcal{G} = \Psi_{\theta_1}(\mathcal{D_L}).
\end{align}

The LCG consists of latent context vectors of dimension $n_{c}$ at each spatio-temporal point $(x,z,t)$,
\begin{align}
    \mathcal{G} = \left\{\left(\mathbf{x}_{1}, \mathbf{c}_{1}\right), \ldots, \left(\mathbf{x}_{M}, \mathbf{c}_{M}\right)\right\}
\end{align}
where $M=n_{1}\times n_{2}\times n_{3} \times n_{c}$ and $n_{1}$, $n_{2}$, and $n_{3}$ correspond to the number of points in the two spatial dimensions ($n_{1}$ and $n_{2}$) and the one time dimension ($n_{3}$).

While the above architecture is standard practice for the application of CNNs in analyzing images, MFFN also includes architectural features for inputing and evaluating the PDEs of the system and generating an additional equation loss function. This ``physics-informed'' extension is characterized by an MLP branch, denoted as the function $\Phi_{\theta_2}$ with trainable parameters $\theta_2$. Incorporation of $\Phi_{\theta_2}$ aims to improve upon the physical plausibility of the final high-resolution prediction. Such physics-informed losses have been widely used in recent years~\citep{RaissPerdikarisKarniadakis2019}. Furthermore, another primary advantage of adopting an MLP in the decoding arm with spatial coordinate inputs is that this approach leads to the continuous nature of the model. The LCG $\mathcal{G}$ can be queried at an arbitrary number of points by performing a trilinear interpolation between the vertices around each query point. In this present problem configuration, there are two space dimensions and one time dimension so each query point has eight neighbors.

In Figure~\ref{fig:architecture}, this random sampling is shown as the small pink box being extracted from the LCG, and the small yellow square in the corner of the extracted box represents a latent context vector at the grid corner. $\mathbf{x}$ and $\mathbf{c}$ are concatenated to form the input array, denoted as the green block in Figure~\ref{fig:architecture}. For each query, the required physical channels of interest are decoded at each of the 8 bounding vertices courtesy of $\mathbf{c}$ and a trilinear interpolation between the centre of the cell and the bounding vertices results in the final value at the query location, which is continuous is nature. 
This input array is also concatenated to the output of the subsequent 4 fully-connected layers in the MLP, which are denoted as the purple blocks in Figure~\ref{fig:architecture}. After each fully-connected layer is a LeakyReLU activation layer.

The final four-dimensional output of this MLP contains the prediction of the values of temperature $\widehat{T}$, pressure $\widehat{P}$, and vertical and horizontal components of velocity $\hat{u}$ and $\hat{v}$, respectively and henceforth denoted as $\widehat{\mathbf{y}}$. These are outputs are represented by the pink-blue blocks in Figure~\ref{fig:architecture}. 
\begin{align}
    \widehat{\mathbf{y}}=\Phi_{\theta_2}(\mathbf{x}, \mathbf{c}), 
\end{align}
where:
\begin{align}
    \widehat{\mathbf{y}} :=
    \lr{\widehat{u}, \widehat{v}, \widehat{P}, \widehat{T}}.
\end{align}
The superresolved, high-resolution prediction $\widehat{\mathbf{y}}$ is compared to the values at the same location in the high resolution DNS 
using the mean-squared-error loss function, also known as an $l_2$ regularized loss:
\begin{align}
  \mathcal{L}_{\mathcal{P}} = \dfrac{1}{N_{B}}\sum_{i=1}^{N_{B}}{\left[\dfrac{1}{N}\sum_{j=1}^{N}{\lr{\mathbf{y}_{j}^{i} - \widehat{\mathbf{y}}_{j}^{i}}\cdot\lr{\mathbf{y}_{j}^{i} - \widehat{\mathbf{y}}_{j}^{i}}}\right]} \label{eq:pred_loss}
\end{align}
where $N_{B}$ is the mini-batch size and $N$ is the number of points for each slab in the training point.

The PDE loss is given by 
\begin{align}
  \mathcal{L}_{\mathcal{E}} = \dfrac{1}{N_{B}}\sum_{i=1}^{N_{B}}{\left[\dfrac{1}{N}\sum_{j=1}^{N}{|\mathcal{R}_{M}\lrb{\widehat{\sol}}}| + |\mathcal{R}_{C}\lrb{\widehat{\sol}}| + |\mathcal{R}_{E}\lrb{\widehat{\sol}}|\right]} \label{eq:eq_loss}
\end{align}

The final loss is a combination of the predictive loss and the equation loss weighted by a constant $\gamma$:
\begin{align}
    \mathcal{L} = \mathcal{L}_{\mathcal{P}} + \gamma \mathcal{L}_{\mathcal{E}},
\end{align}
where $\gamma$ was determined to be $\gamma=0.05$ in J20. We adopt $\gamma=0.05$ for the remainder of the paper. 

\begin{table}[ht]
\caption{table of hyperparameters for NNs utilized}
\centering
\begin{tabular}{l|l|l}
    \hline
    Hyperparameter      & UNet               & MLP\\
    \hline
    Layers              & 3 (UNet Layers)    & 6 (Dense Layers) \\
    Activation function & ReLU               & LeakyReLU        \\
    Stride              & 1                  &                  \\
    Kernel              & (1,3,1)            &                  \\
    Input features      & 4                  & 32               \\
    Output features     & 32                 & 4                \\ 
    \hline
\end{tabular} \label{tab:hyperparameters}
\end{table}

A summary of some of the hyperparameters of the NNs utilized in the architecture employed is summarized in Table~\ref{tab:hyperparameters}. Further details on the network architecture can be found in the original J20 work and its implementation can be found on their GitHub repository: \href{https://github.com/maxjiang93/space_time_pde/tree/master}{https://github.com/maxjiang93/space\_time\_pde/tree/master}.

    
\subsection{This study's architectural modifications}

In this work we add two additional physical constraints to the pipeline. The first additional physical constraint is the inclusion of RBC systems' full set of PDEs in determining $\mathcal{L}_{\mathcal{E}}$. The original J20 implementation neglected the inclusion of the mass continuity Equation~\eqref{eq:cont} in considering their PDE loss. We include this term in order for the loss to reflect the full set of the system's PDEs. This modification is reflected in the second to last term in~\eqref{eq:eq_loss}. We further implement physical constraints by introducing a boundary loss, which is a standard implementation in many PINNs architectures \citep{RaissPerdikarisKarniadakis2019, BerroneEtAl2023}, and helps to satisfy the required boundary conditions. In this particular geometry for RBC, periodicity along the x-axis and constant temperatures on the top and bottom plates in the model-predicted output are enforced. We denote this boundary loss as $\mathcal{L}_{\mathcal{B}}$, and it mirrors the procedure of the prediction loss except that it only compares the values on the boundaries:
\begin{align}
  \mathcal{L}_{\mathcal{B}} = \dfrac{1}{N_{B}}\sum_{i=1}^{N_{B}}{\left[\dfrac{1}{N_{b}}\sum_{j=1}^{N_{b}}{\lr{\mathbf{y}_{j}^{i} - \widehat{\mathbf{y}}_{j}^{i}}\cdot\lr{\mathbf{y}_{j}^{i} - \widehat{\mathbf{y}}_{j}^{i}}}\right]} \label{eq:boundary_loss}
\end{align}

where $N_{b}$ is the number of boundary points and the sum is understood to be only over the points on the boundary.

When considering this additional loss, the resultant loss is the total sum of the predictive loss, the equation loss, and the boundary loss:
\begin{align}
     \mathcal{L} = \mathcal{L}_{\mathcal{P}} + \gamma\lr{\mathcal{L}_{\mathcal{E}} + \mathcal{L}_{\mathcal{B}}}. \label{eq:loss}
\end{align}
Note that in the current work the equation loss and boundary loss use the same weighting factor $\gamma$.

\section{Results} \label{sec:results}

\subsection{Training results}
 \begin{figure*}
     \includegraphics[width=\linewidth]{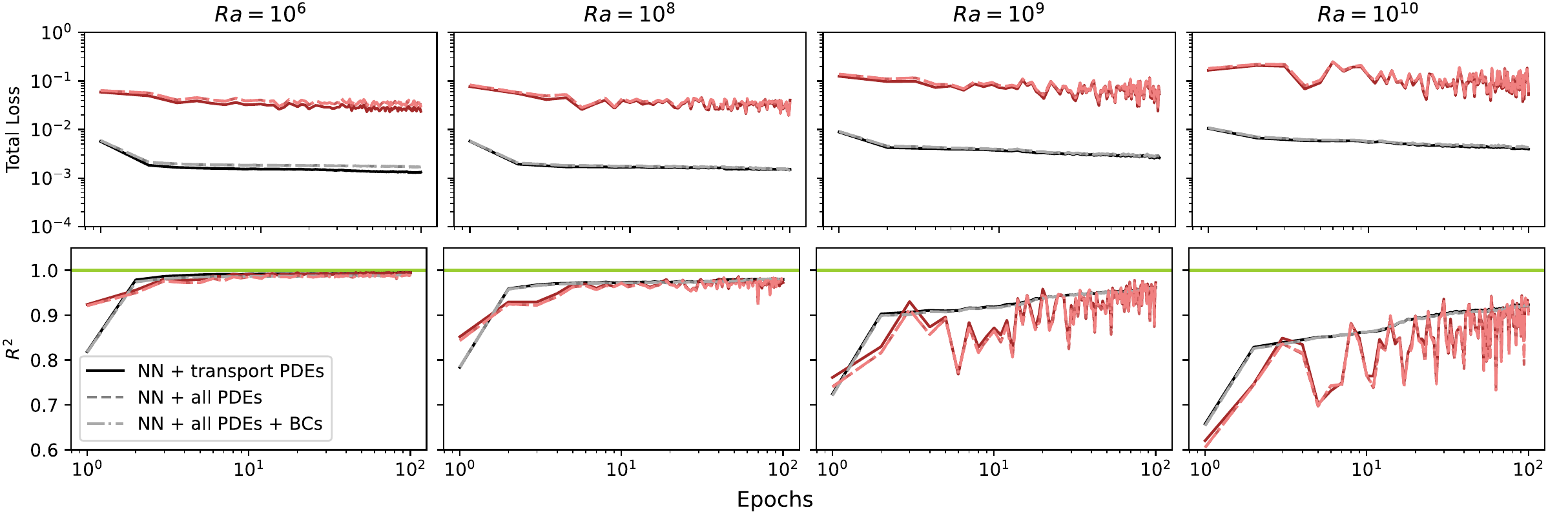} 
     \caption{Evolution of loss values (\textit{top row}) and goodness-of-fit coefficient ($R^2$) (\textit{bottom row}) by training epoch for $Ra=10^6$, $10^8$, $10^9$ and $10^{10}$, for the case of $\gamma=0.05$. An $R^2$ of 1 indicates a perfect fit, which is shown as the solid green line in the bottom row panels. The black/grey gradient and red gradient sets of lines show the curves for the training and validation sets respectively. The solid lines show the experiments run on the original J20 architecture which includes the NN and transport PDEs contributing to the equation loss, the dashed lines showing experiments with the addition of the continuity equation contributing to the equation loss and the dash-dotted lines showing experiments implementing both the full set of PDEs in the equation loss as well as the boundary loss.} \label{fig:SAMPLE_LOSS_R2_CURVE}
 \end{figure*} 

We present the results of training our adopted neural networks for 100 epochs, conducted for each of the simulations considered of $Ra=10^6$, $10^8$, $10^9$ and $10^{10}$, with an Adam optimizer and learning rate of $10^{-2}$. We choose 100 epochs to stay consistent with the \citet{Jiang2020} study's training parameters. The training runs for each $Ra$ are conducted independently, such that the training of one $Ra$ value has no influence on that of the other $Ra$ values explored in this study. Figure~\ref{fig:SAMPLE_LOSS_R2_CURVE} shows the evolution by training epoch of the training (black/grey gradient) and validation (red gradient) losses as defined by \eqref{eq:loss} on the top row, as well as the $R^2$ evolution on the bottom row. 

The $R^2$ is called the coefficient of determination and is a measure of goodness-of-fit, i.e., the degree to which the observed outcomes (in our case, the field values of super-resolved predictions) match that of the theoretical model (being the original DNS simulation here), with a perfect fit being indicated by an $R^2$ value of $1$. This is shown as the solid green line at $R^2=1$ in the bottom panels of Figure~\ref{fig:SAMPLE_LOSS_R2_CURVE}.  

We observe that whilst there exists large discrepancies between training and validation losses of about two orders of magnitude, final $R^2$ values converge to 1. We do not observe an inflection point in validation loss across any $Ra$ experiment, which suggests that the network is not over-fitting. However, we observe that the validation curves across all metrics are less stable at higher Rayleigh numbers. Higher Rayleigh number simulations exhibit fields with enhanced turbulence and mode interaction and thus exhibit more complex features. It is therefore not unreasonable that the increasing complexity in more turbulent flows results in an increased difficulty in replicating the absolute values of the DNS fields. We also observe that training with the inclusion of the full set of PDEs and boundary loss produced similar results as training with only images and the transport equations. This is most visibly illustrated in Figure~\ref{fig:SAMPLE_LOSS_R2_CURVE}, where we see that the loss curves for training and validation respectively all follow a very similar trajectory, regardless of having been trained on the original J20 architecture (which includes the NN and \textit{only} the transport equations), CNN and full set of PDEs (i.e., inclusive of all transport equations \textit{and} the continuity equation) or CNN, full set of PDEs and boundary loss, resulting in the curves stacking on top of each other in the figure. This suggests that even with additional physical constraints, the main factor driving the network's prediction for this physical setup, geometry and set of  hyperparameters, is the spatial (image) features of the simulation. 

To analyse the performance of evaluating our fully trained models, we coarsen the native resolution of each of the full DNS runs in our suite by a factor of 4, and pass it through the architecture described in Section~\ref{sec:methods} in inference mode, i.e., with the architecture loaded with the weights and biases saved from training, to attain SR predictions of the same dimensions as the native DNS runs. The following discussions will refer to the results of these predictions attained in inference mode. 

In Figures~\ref{fig:SR_figureRa6}-\ref{fig:SR_figureRa10}, we showcase a comparison of a snapshot of the DNS run's $u$-velocity, $v$-velocity and temperature fields and the corresponding super-resolved predictions of these fields. In the one non-turbulent regime at $Ra=10^6$ in this investigation, shown in Figure~\ref{fig:SR_figureRa6}, we see that there are no small-scale or high-frequency structures in the native DNS; only the large oscillatory modes. The SR prediction does an excellent job at reproducing these large modes.  However, at the very small scales, the super-resolved prediction results in the generation of systematic, rectangular, high-frequency features. The energies at the small scales of the SR prediction would therefore contain more energy than that of the DNS at the same scales and this would in turn make the SR prediction \textit{less} dissipative than the DNS. This kind of artifact is often observed in networks which exhibit architectures that include deconvolution layers \citep{Salimans2016, Donahue2016, Radford2015, Dumoulin2016} and is known as the ``checkerboard effect''. We further highlight this effect when discussing the power spectrum later in this section.

We see that qualitatively, smaller and smaller structures begin to appear in the DNS as we get to increasingly high Rayleigh numbers as shown in Figures~\ref{fig:SR_figureRa8},\ref{fig:SR_figureRa9} and \ref{fig:SR_figureRa10}. 
Again, the SR does an excellent job at reproducing the large and medium scale features of the flow. However, the small scales predicted by the SR are somewhat smeared out. This indicates that the SR predictions are more dissipative than their DNS counterparts in these regimes.
 
\subsection{Power Spectra Analysis}

 \begin{figure*} 
     \centering
     \includegraphics[width=\linewidth]{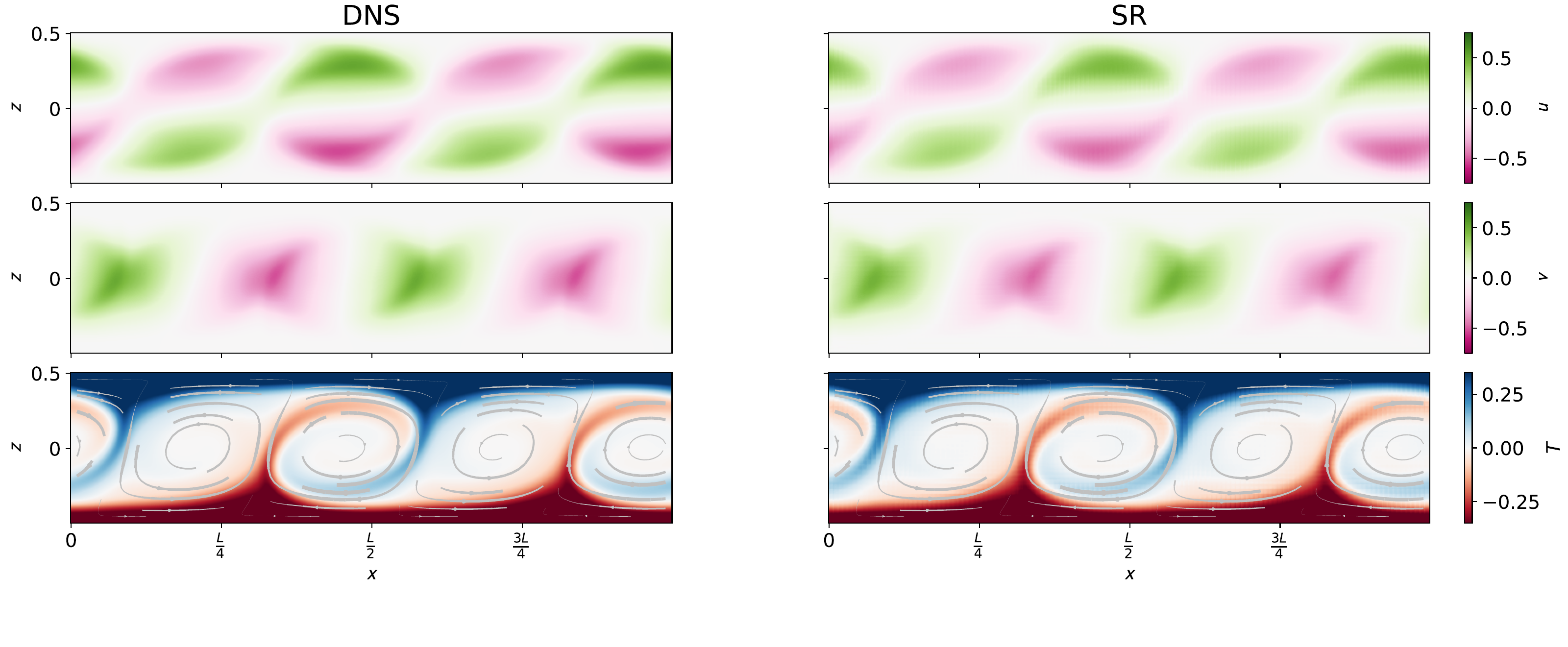}
     \caption{Output super-resolved prediction from CNN (right columns) compared to ground truth simulations (left columns) for $u$-velocity (top rows), $v$-velocity (middle rows) and temperature fields (bottom rows, streamlines overlaid) for non-turbulent RBC of $Ra=10^6$.} \label{fig:SR_figureRa6}
 \end{figure*}

 \begin{figure*}
     \centering
     \includegraphics[width=\linewidth]{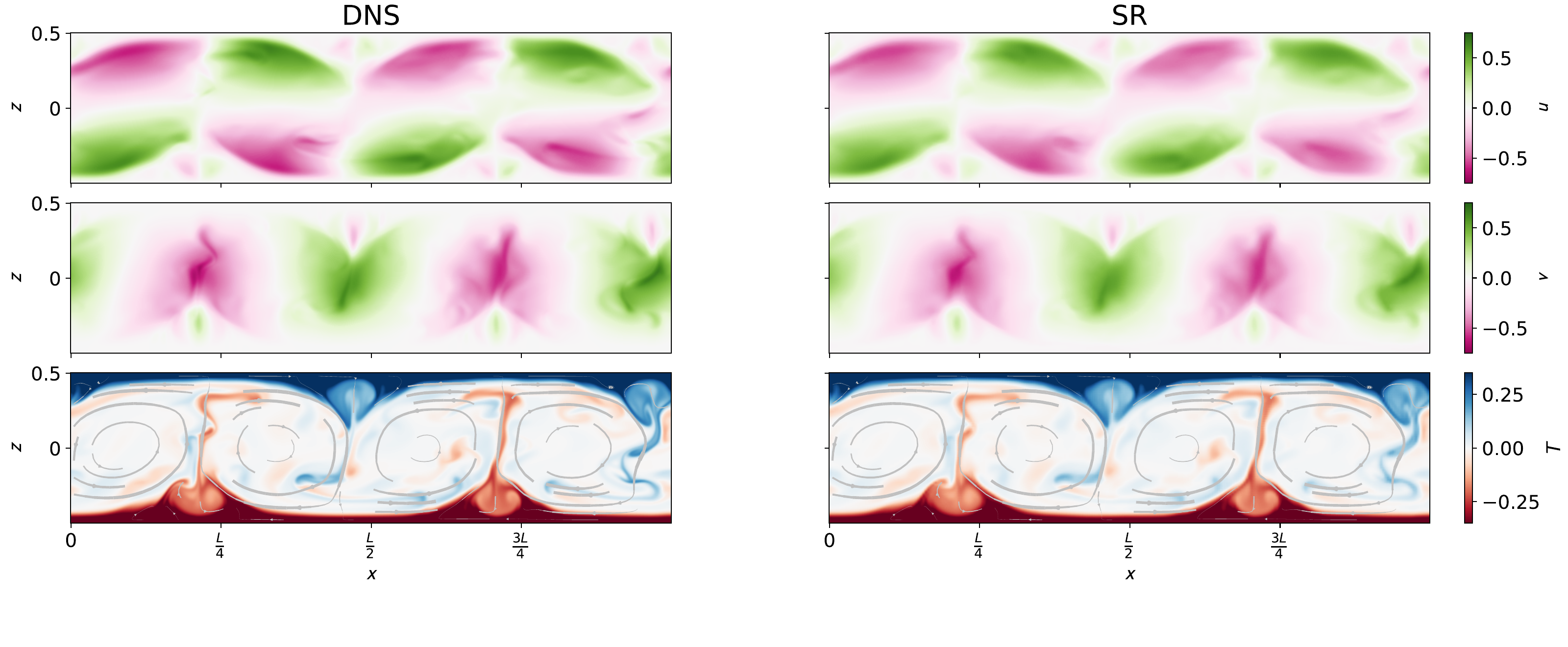}
     \caption{Same as Figure~\ref{fig:SR_figureRa6}, for turbulent RBC of $Ra=10^8$.}
     \label{fig:SR_figureRa8}
 \end{figure*}

 \begin{figure*}
     \centering
     \includegraphics[width=\linewidth]{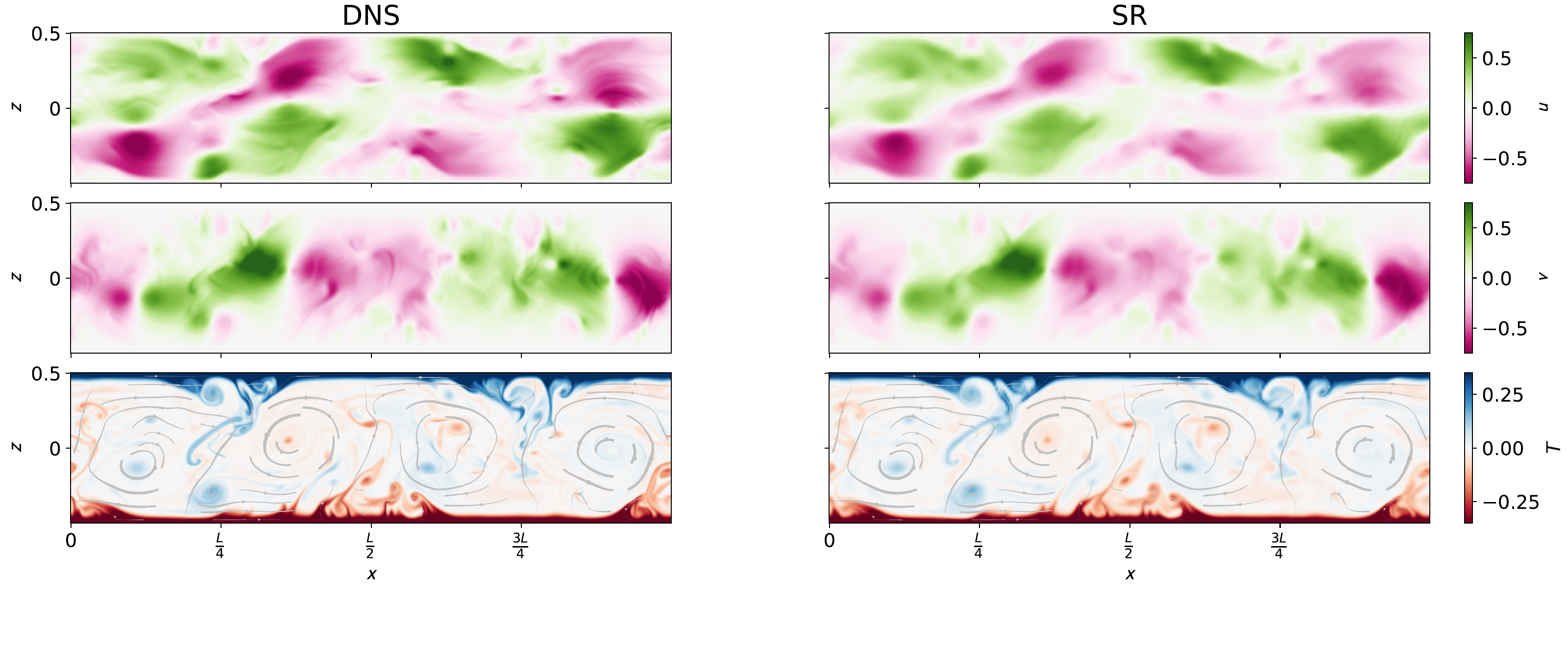}
     \caption{Same as Figure~\ref{fig:SR_figureRa6}, for turbulent RBC of $Ra=10^9$.}
     \label{fig:SR_figureRa9}
 \end{figure*}

 \begin{figure*}
     \centering
     \includegraphics[width=\linewidth]{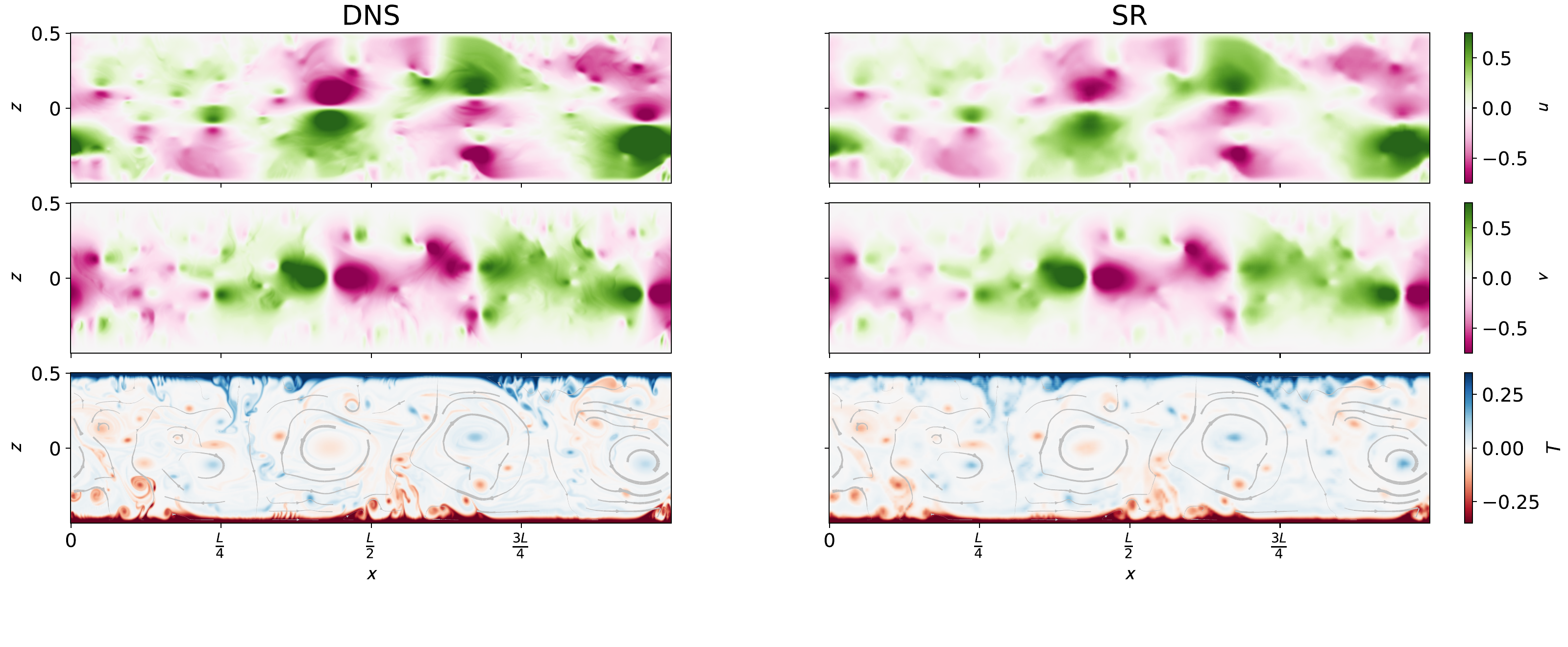}
     \caption{Same as Figure~\ref{fig:SR_figureRa6}, for turbulent RBC of $Ra=10^{10}$.}
     \label{fig:SR_figureRa10}
 \end{figure*}

\begin{figure*}
     \centering
     \includegraphics[width=0.9\linewidth]{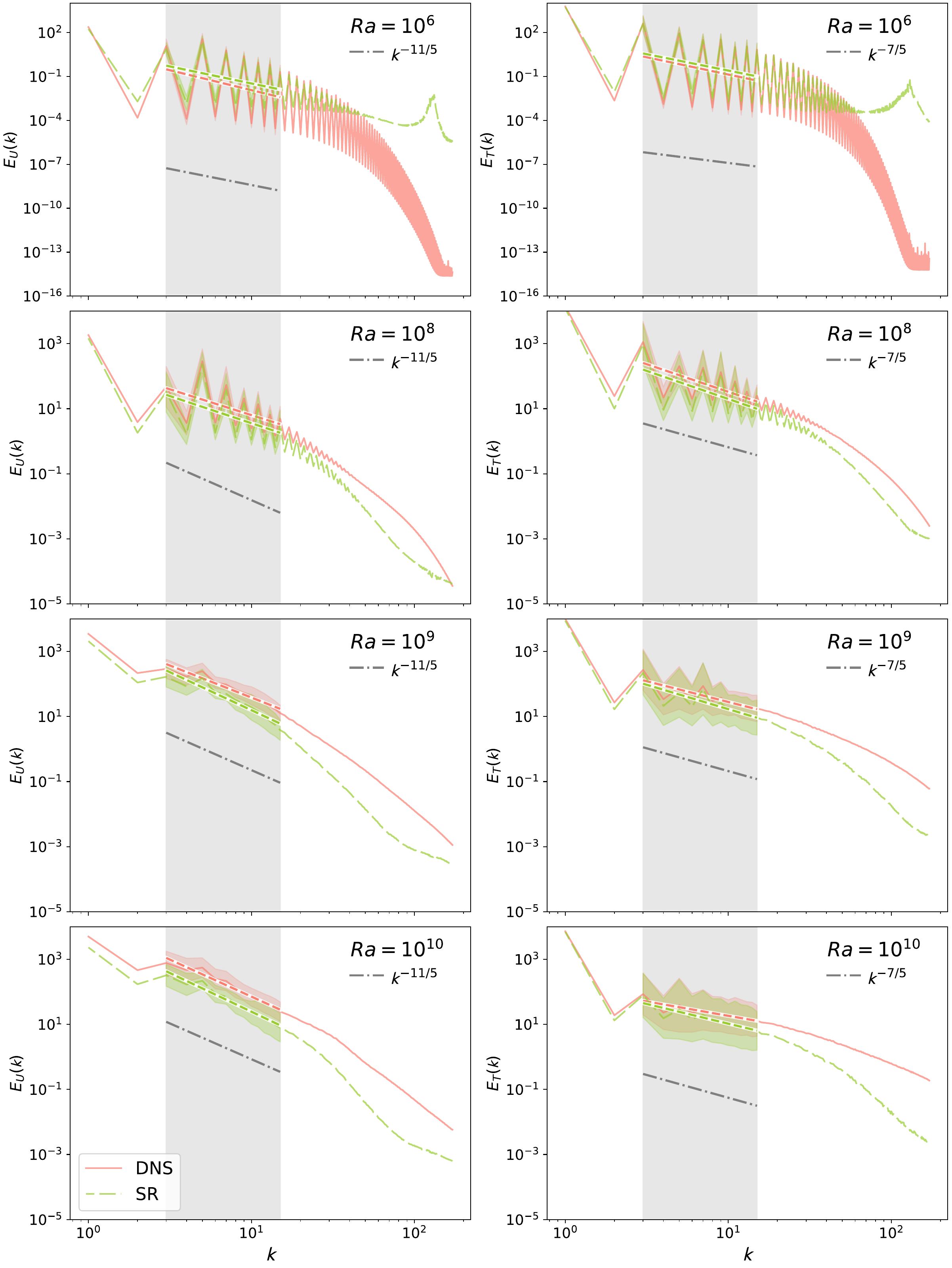}
     \caption{The averaged-in-time power spectra of the total kinetic energy derived from the $u$ and $v$-velocities \textit{(left columns)} and temperature \textit{(right columns)} fields of the DNS (pink solid lines) and SR prediction (green long-dashed lines) for $Ra=10^6-10^{10}$ (\textit{top to bottom panels, in increasing order}). The shaded pink and green regions are the standard deviations from the mean for the DNS and SR prediction respectively. We show the inertial range of $k=3-15$ as the shaded region in grey.
     For this region, we show the best linear fit to the power spectra in the short-dashed overlaid lines for the data, and the offset dark dot-dashed grey line shows the theoretical slope.
     }
     \label{fig:power_spectra}
 \end{figure*}

To quantify the degree to which the SR network described in Section~\ref{sec:methods} captures the physical properties of the DNS, we employ the power spectrum; a two-point statistical tool to determine how energy is distributed as a function of scale. 
In Figure~\ref{fig:power_spectra}, shows energy spectra for slices along the streamwise direction averaged over the wall-normal coordinate and time, which we refer to as the averaged energy spectra. The averaged total kinetic energy spectrum and temperature spectrum are denoted by $E_U(k)$ and $E_T(k)$, respectively. 
Figure~\ref{fig:power_spectra} depicts comparisons of the DNS and SR energy spectra for each of the Rayleigh numbers investigated in this study, in increasing order from top to bottom. The standard deviation in time  
is shown in the shaded areas above and below the lines. 
We also show the linear fit of the spectra within the approximate inertial range of $k=3$-$15$ 
to quantify the degree to which the slope deviates from that of the DNS and the analytic theoretical slopes of $k^{-11/5}$ for the total kinetic energy, and $k^{-7/5}$ for the temperature fields for 2D RBC \citep{Koolath2020}. 
We choose the inertial range of $k=3$-$15$ because this is the range after which dissipation is observed for the fluid simulations presented in other works~\citet{Koolath2020, BurkhartLazarian2015}. 

We make the following observations regarding these power spectra. Firstly, we take note that the oscillatory nature of the low-$Ra$ flows is seen very prominently. The $Ra=10^6$ and $10^8$ power spectra show clear systematic fluctuations within the chosen inertial range, and these oscillations continue for the entire range of $k$ values in the $10^6$ case, showing that this system might not really exhibit a proper inertial range. Said oscillations continue to about $k=50$ in the $10^8$ case. Oscillatory fluctuations signaling the presence of the largest scale modes are observed consistently across the power spectra of all $Ra$ cases in the \textit{lowest} $k$ values, but the power spectra in the chosen inertial range and dissipation scales are visually much smoother in the $10^9$ and $10^{10}$ cases, providing clear evidence of the turbulent nature of the flows in these high $Ra$ simulations. This has important implications to consider because the theoretical slopes shown in the dash-dot grey lines underneath the power spectra in Figure~\ref{fig:power_spectra} are calibrated against \textit{turbulent} flows, thus a comparison to theory becomes more appropriate as we get to the higher $Ra$ cases. 

Turning our attention again to the top row of Figure~\ref{fig:power_spectra}, which showcases the results of the non-turbulent, $Ra=10^6$ regime, we notice that the aforementioned artificial high-frequency feature spatially manifesting as systematic, grid-like structures in Figure~\ref{fig:SR_figureRa6} can be clearly observed in the spike in the high-$k$ energies in both the total kinetic energy and temperature fields' spectra, signaling that this is a nonphysical feature and instead an artifact of the NN pipeline described in Section~\ref{sec:methods}. 

In contrast, the energy spectra of the super-resolved fields in the turbulent ($Ra=10^{8}-10^{10}$) regimes approximately mirror the shape of the DNS spectra more closely, albeit at lower energies which signals a more dissipative field. This suggests that although the energies are not captured faithfully to the smallest scales, which is evident in the smoothing out of small-scale features seen in Figures~\ref{fig:SR_figureRa8} and~\ref{fig:SR_figureRa10}, the resultant field is more physically realistic than that produced by the non-turbulent case. 

The above observations and results are generally consistent between the $E_U(k)$ and $E_T(k)$, but especially at the smallest scales at high $Ra$, the power spectra between the DNS and SR predictions seems to agree slightly better in $E_U(k)$. We also performed the above-described experiment with disregard for the PDE and Boundary Losses, i.e., for the case of $\gamma=0$, and found that there exhibited little to no difference in the presentation of the power spectrum of the predicted fields, as is the case with the training curves in Figure~\ref{fig:SAMPLE_LOSS_R2_CURVE}. Whilst this may naively suggest that the PDE evaluation component of the architecture presented in Figure~\ref{fig:architecture} is redundant, we acknowledge the lack of hyperparameter tuning especially at the high $Ra$ experiments, to which training and subsequent results may be very sensitive. We discuss this point futher in the limitations and caveats component of Section~\ref{sec:discussion}. These architectural features, as well as $\mathcal{L}_\mathcal{E}$ and $\mathcal{L}_\mathcal{B}$, should therefore not be completely ruled out in future works and development of PINNs.

As these above observations would not have been obvious from a judgement based purely on the final loss and $R^2$ values, we have demonstrated the critical value of employing the power spectrum to critique the degree to which an ML-learnt turbulent field is physically realistic. 

\begin{figure}
     \centering
     \includegraphics[width=\linewidth]{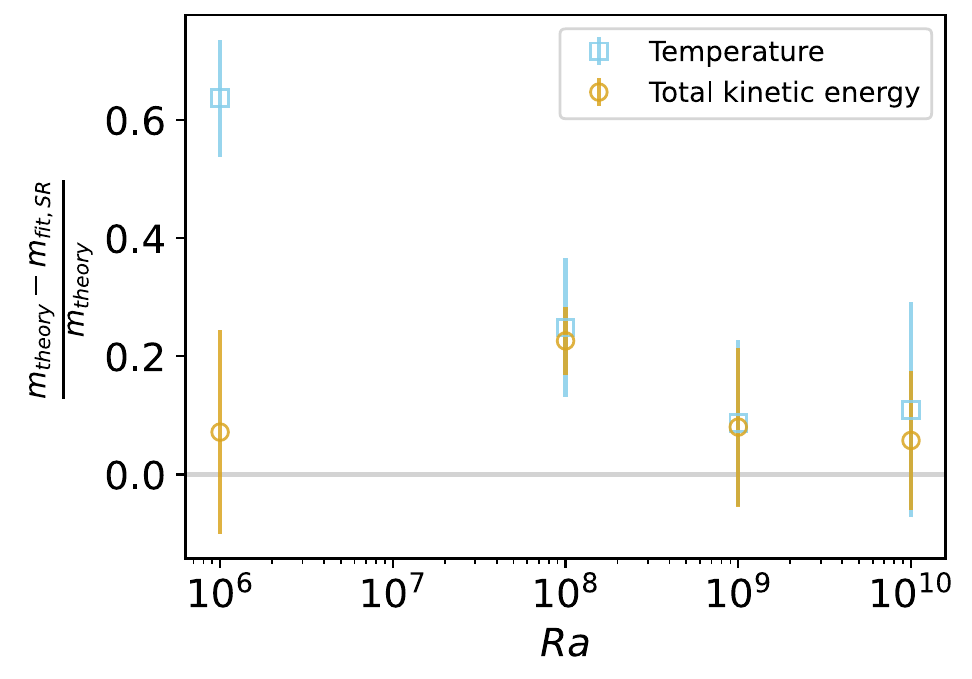}
     \caption{Relative error of the best-fit linear slopes of the power spectrum of the super-resolved fields' total kinetic energy (\textit{yellow circles}) and temperature (\textit{blue squares}), in the inertial range ($k=3-15$), normalized to the theoretical slopes.}\label{fig:residuals}
 \end{figure}

To investigate the degree to which the absolute value of the line of best fit to the SR field power spectra in the inertial range deviate from analytical theory, we calculate the relative error between the time-averaged fit and the theoretical slope. 
The results from the cases considered are shown in Figure~\ref{fig:residuals}. A value of $0$ indicates a perfect agreement between the SR fields' power spectra's slope of best fit and the theoretical values of $m_{\mathrm{theory}}=-11/5$ for the total kinetic energy and $m_{\mathrm{theory}}=-7/5$ for the temperature fields. Therefore, larger deviations from $0$ would signify a greater departure from a physically realistic field. The error bars in Figure~\ref{fig:residuals} indicate the standard deviation in time of the relative error, so we may deem points with error bars that touch $0$ to have sufficient agreement between the power law slope of the power spectrum and that of theory. 

Whilst most relative error values of the total kinetic energy field represented by the yellow circles in Figure~\ref{fig:residuals} lies around $0$ within uncertainty, there is a clear trend demonstrating that the relative error of the temperature field, represented by the blue squares in Figure~\ref{fig:residuals}, decreases with increasing $Ra$, with the non-turbulent case at $Ra=10^6$ showing the greatest deviation from $0$. This deviation decreases with increasing $Ra$. At the highest $Ra$ of $10^9$ and $10^{10}$, a deviation of $0$ falls within the uncertainty of time variations. 

From these analyses, it is evident that the NN-based pipeline that had been given physics-informed priors, is more adept at describing energy and temperature spectra on most scales in more turbulent systems, whereas the full spectrum of non-turbulent or transient fluid flows are not efficiently captured by this method. This result testifies to the ML pipeline's ability to recreate the statistics in the inertial range of a system that exhibits a broad range of spatial frequencies. However, we also emphasize that whilst these results for the inertial range appear promising, upon re-examination of Figure~\ref{fig:power_spectra} it is clear that these slopes start deviating from theoretical slopes and DNS power spectra rapidly as one starts to consider the smaller, dissipative scales. We therefore reiterate our conclusion that the method presented in Section~\ref{sec:methods} is adequate for super-resolution to capture large scale structures in the inertial range for turbulent systems, but does not reliably capture fine-scale structures.

\section{Discussion} \label{sec:discussion}
Despite the successes of numerical simulations of turbulence in reproducing both analytic scaling predictions  \citep{Kolmogorov1991, GoldreichSridhar1995} and the observed turbulent density and velocity power spectral scaling relations \citep{Armstrong1995,2022arXiv220413760Y}, a number of major  practical challenges remain.  Critically, the physical resolution of numerical simulations is severely limited, restricting research to a narrow range of scales to study the turbulence inertial range. 
Failing to resolve turbulent systems to the smallest relevant scales would, for example, underestimate the amount of energy within the system~\citep{Jin2017} or misdiagnose the power-law slope of the energy spectrum. As different power law slopes are explained by different theoretical frameworks, resolution becomes critical for distinguishing theories of turbulence. Current high-resolution turbulence simulations can achieve $Re \approx 10^{5}$, running on more than $65,000$ cores \citep{Federrath2016, Federrath2021}, and such simulations remain remote from realistic values of up to $Re=10^{10}$ for the turbulent interstellar medium.

Superresolution ML studies may be a promising avenue for achieving high resolution numerical simulations at a fraction of the computational cost. In this study, we have presented a modified MeshFreeFlowNet architecture that integrates the complete set of simulation PDEs and boundary conditions to enhance the resolution of numerical simulations of turbulent RBC. Our approach is an experimental step towards addressing the longstanding challenge of limited resolution in astrophysical turbulence simulations by employing ML techniques for superresolution. We find that the model successfully recovers large-scale information and the inertial range scaling in both convective and highly turbulent RBC systems, with a training set encompassing $Ra$ ranging from $Ra=10^6$ to $10^{10}$.

We have demonstrated that the superresolution architecture explored here is successful at reproducing the statistics of more turbulent flows, where a rich variety of structures exists at multiple length scales. This is in contrast to laminar flows, which exhibit a lower degree of complexity. Furthermore, our analysis also indicates that since higher Rayleigh numbers exhibit richer features within fewer free-fall times the CNN is able to capture more relevant information. This finding suggests that this model is particularly well-suited for studying highly turbulent systems and could prove valuable in advancing the achievable resolution of astrophysical turbulence simulations.  

In order to assess the model's performance in recreating the full range of spatial frequencies present in the ground truth simulation, we compared the superresolution's predicted power spectrum slopes to the theoretically expected power spectrum slope within the inertial range. Our analysis revealed that both the total energy and temperature power spectrum are well-recovered and agree well with theory within our chosen inertial range of $k=3-15$ for the turbulent $Ra=10^9$ and $Ra=10^{10}$ simulations. Turbulent systems exhibit mode-mode interactions which creates a hierarchy of interacting phases \citep{Peek2019}, leading to richer phase structure to train on. We therefore speculate that the procedure performs well on more turbulent systems as the NN can better extract information and fill in at the small scales. These results demonstrate the effectiveness of our modified MeshFreeFlowNet in capturing the essential features of RBC systems for turbulent simulations. 

On the other hand, we have observed high-frequency features in low $Ra$ simulations, which result in an artificial energy pile-up, or in the ML community, a ``checkerboard effect''. This phenomenon is an unphysical systematic artifact of the chosen architecture and may impact the accuracy of the model in lower $Ra$ regimes. Further investigation into the causes of this energy pile-up and possible mitigation strategies will improve the model's performance across a broader range of $Ra$ values. 

Despite the promising results, there are potential avenues for further improvement. Future research could focus on incorporating additional physics constraints for turbulent systems, such as incorporating anisotropy or intermittency, to enhance the model's ability to capture the complex dynamics of turbulent systems. The model validation could also include higher-order statistics~\citep{Burkhart2021} in addition to our use of the power spectrum in this work. 
Additionally, the model could be extended to study other types of astrophysical turbulence, such as magnetohydrodynamic (MHD) turbulence~\citep{guan2023mhdnet}, which would require altering the PDEs of the system and retraining on a simulation suite such as the Catalog for Astrophysical Turbulence Simulations (CATS; \citet{Burkhart2020}).

Furthermore, we acknowledge the following caveats and limitations of this study that need be considered:

The performance of our model is sensitive to the choice of the hyperparameter $\gamma$, which may affect the balance between the physics constraints and the neural network's ability to learn complex features. Future work should explore techniques for optimizing $\gamma$, such as grid search or Bayesian optimization, together with proper cross validation to improve model performance across a range of Rayleigh numbers.

The quantity and quality of the training data, particularly the number of turnover times captured, can have a significant impact on the model's ability to learn the underlying physics of the system. Further research should investigate the relationship between the number of turnover times and the model's performance, with a focus on determining the optimal amount of training data required for accurate turbulence prediction.

The current model is limited to 2D simulations of RBC systems. However, astrophysical turbulence phenomena occur in three dimensions, necessitating the extension of the model to 3D simulations. This transition will likely require further modifications to the neural network architecture, as well as additional computational resources to accommodate the increased complexity of 3D systems.

Another method to increase the accuracy of the model may be to add in an additional MLP branch or other architectural features to consider known numerical metadata such as the Rayleigh number of the ground truth simulations. Such considerations should be taken into account when optimising for models in future works.
    
A deficiency in the model adopted in this study is the requirement for the input spatial coordinates and an MLP in the decoding arm (mapping from the latent context grid to the high resolution input) of the model does not guarantee equivariance with respect to the low resolution inputs, although it is indeed what facilitates the advantageous continuous nature of the model. The current model could be endowed with approximate equivariance properties via data augmentation such as those presented in \citet{LingJonesTempleton2016} or alternatively, a new architecture, but these are out of the scope of the present study.

We have observed that the model tested in this study does not accurately capture the small scale features of the ground truth turbulence simulations. This may be due to the deterministic nature of incorporating a NN with a mean squared error loss. Future works could consider integrate generative models that models the posterior of high resolution simulations given a low resolution one, for instance. Architectural features that have been shown to perform well on reconstruction and learned turbulence simulation, such as dilated CNNs \citep{Stachenfeld2021, YuKoltun2016} could also be considered. Finally, to address the "checkerboard effect" one could consider architectural features like substituting nearest-neighbour upsampling with bilinear interpolation in the deconvolution layers, or by inserting a fixed convolutional layer after each upsampling layer and before each downsampling layer \citep{KinoshitaKiya2020}.

In conclusion, while our modified MeshFreeFlowNet architecture has shown promise for superresolution turbulence studies, addressing the aforementioned caveats and limitations will be crucial for further improving the model's performance and applicability to a broader range of astrophysical systems.

\section{Conclusions} \label{sec:conclusions}
In this work, we have modified the MeshFreeFlowNet architecture to include the full set of simulation PDEs for an RBC system as well as a loss function for the simulation  boundary conditions. We study and contribute an independent measure of the network's performance and ability to recreate physically realistic fields on a range of Rayleigh numbers from $Ra=10^6$-$10^{10}$ in the statistically steady regime by leveraging the statistical analysis tools of turbulent fluids. By doing so we find that:

\begin{itemize}
    \item The network performs well at all Rayleigh numbers studied here in recovering large-scale spatial features and the power spectrum at small wave numbers.  
    \item We find that our updated architecture performs better at capturing the small-scale features and power spectrum in more turbulent numerical setups (i.e., larger $Ra$ systems) than that of oscillatory flow systems. This is likely due to the fact that more turbulent systems have a rich variety of structures at many length scales in comparison to laminar flows.
     \item  For $Ra>10^{8}$, the superresolution CNN network used here is overly dissipative at large wavenumbers compared to the direct numerical simulations but still recovers the theoretical slope of the inertial range for both temperature and kinetic energy fields. 
\end{itemize}
 \acknowledgements

D.M.S. is supported by the 2022 Future Investigators in NASA Earth and Space Science and Technology (NASA FINESST) Fellowship (NASA grant 80NSSC22K1604). D.M.S. is also grateful for the generous support of the 2023 Quad Fellowship.  
 B.B. acknowledge support from NSF grant AST-2009679 and NASA grant No. 80NSSC20K0500.
B.B. is grateful for the generous support of the David and Lucile Packard Foundation and the Alfred P. Sloan Foundation. 
The Flatiron Institute is supported by
the Simons Foundation. All authors are grateful for the use of both GPU and CPU resources on the Rusty Supercomputer cluster at the Simons Foundation Flatiron Institute.

\bibliographystyle{aasjournal.bst}
\bibliography{super_res}
\label{lastpage}

\end{document}